\documentclass{emulateapj}

\usepackage{graphicx}
\usepackage[dvipsnames]{color}

\newcommand{\bl}[1]{\mbox{\boldmath$ #1 $}}
\newcommand{\beq}{\begin{equation}}
\newcommand{\eeq}{\end{equation}}
\newcommand{\barray}{\begin{eqnarray}}
\newcommand{\earray}{\end{eqnarray}}
\newcommand{\bfig}{\begin{figure}}
\newcommand{\efig}{\end{figure}}

\shorttitle{The Burst Mode of Accretion}
\shortauthors{Vorobyov et al.}

\begin{document}

\title{The Burst Mode of Accretion in Primordial Protostars}

\author{Eduard I. Vorobyov\altaffilmark{1,}\altaffilmark{2}, Alexander L. DeSouza\altaffilmark{3}, Shantanu Basu\altaffilmark{3}}

\altaffiltext{1}{University of Vienna, Institute of Astrophysics,  Vienna, 1180, Austria; eduard.vorobiev@univie.ac.at.} 
\altaffiltext{2}{Research Institute of Physics, Southern Federal University, Stachki 194, Rostov-on-Don, 344090, Russia.}
\altaffiltext{3}{Department of Physics and Astronomy, University of Western Ontario, London, Ontario, N6A 3K7, Canada; alexander.desouza@gmail.com, basu@uwo.ca.}

\begin{abstract}
We study the formation and long-term evolution of primordial protostellar disks harbored by first stars using numerical hydrodynamics simulations in the thin-disk limit. The initial conditions are specified by pre-stellar cores with distinct mass, angular momentum, and temperature. This allows us to probe several tens of thousand years of the disk's initial evolution, during which we observe multiple episodes of fragmentation leading to the formation of gravitationally bound gaseous clumps within spiral arms. These fragments are torqued inward due to gravitational interaction with the spiral arms on timescales of $10^3$--$10^4$~yr and accreted onto the growing protostar, giving rise to accretion and luminosity bursts. The burst phenomenon is fueled by continuing accretion of material falling onto the disk from the collapsing parent core, which replenishes the mass lost by the disk due to accretion, and triggers repetitive episodes of disk fragmentation. We show that the burst phenomenon is expected to occur for a wide spectrum of initial conditions in primordial pre-stellar cores and speculate on how the intense luminosities (${\sim}10^7~\mbox{L}_{\odot}$) produced by this mechanism may have important consequences for the disk evolution and subsequent growth of the protostar.
\end{abstract}


\section{Introduction}
\label{intro}

Cosmological-scale simulations of collapsing primordial clouds in the early universe have been used to suggest that the first luminous objects in the universe were stars with masses of $M\,{\ga}\,100~\mbox{M}_\odot$ that formed in relative isolation \citep[e.g.,][]{Abel2000,Bromm2004,Yoshida2008}. However, the above numerical simulations were effectively based on a scenario of monolithic quasi-spherical collapse of cloud cores. Angular momentum, disk formation, and fragmentation have added greater depth to this picture, replacing it with one in which the collapsing primordial cores of the early universe produce rich structure
in the inner regions where disks emerge.

Although protostellar disks are a ubiquitous outcome of the present-day star formation process, the importance of these structures to the evolution of the first stars has begun to be understood only recently. \citet{Saigo2004} and \citet{Machida2008} have demonstrated that disk-like structures are expected to form from primordial cores with a wide range of initial rotation rates, and can fragment to yield binary pairs of first stars. \citet{Clark2011} were able to follow the collapse of primordial clouds to protostellar densities and found vigourous gravitational fragmentation in primordial disks leading to the formation of tightly bound multiple stellar systems. Using smoothed particle hydrodynamics simulations combined with the sink particle technique, \citet{Smith2011,Smith2012} studied the importance of accretion luminosity on disk fragmentation and the effect of protostellar accretion on the structure and evolution of protostars. \citet{Greif2012} most recently performed high-resolution, three-dimensional hydrodynamic simulations using a Lagrangian moving mesh and found rapid migration and merging of secondary fragments with the primary protostar.


The general picture to emerge from these simulations is that protostellar accretion may be a highly variable process fuelled by disk gravitational instability and fragmentation, in a similar manner to present-day star formation \citep{VB06,VB10,Machida2011}. However, a major problem of high-resolution three-dimensional studies has been the difficulty of following the evolution of primordial protostellar disks for longer than a few thousand years.
In this study we present gas hydrodynamics simulations of collapsing primordial cores into a disk formation phase using the thin-disk approximation. We adopt a barotropic equation of state derived by \citet{Omukai05} for the primordial chemical composition of gas. The benefit of this type of formulation is that we are able to study the fragmentation, evolution and subsequent accretion of protostellar mass clumps formed within the disk, while simultaneously maintaining resolution of the extended remnant parent cloud core over many orbital periods and model realizations.

The structure of this article is as follows. We briefly describe the numerical simulations in Section 2, including the modifications that have been made pertaining to the primordial star-forming environment. In Section 3 we present characteristics of the temporal evolution of our reference model, starting from the prestellar phase, and ending when the mass of the protostar has reached $45~\mbox{M}_{\odot}$. Section 3 also includes an analysis of the protostellar accretion luminosity expected to arise from the burst mode of accretion that is unique to our model of primordial star formation. Finally, in Section 4 we extend our discussion and draw conclusions from these results.

\section{Model description}
\label{model}

Our model and method of solution update the model presented in \citet{VB05,VB06}, with appropriate modifications for star and disk formation in the early universe. We follow the evolution of gravitationally unstable primordial cores from the isolated pre-stellar stage into the protostar and disk formation stage and terminate our simulations once about 30\% of the initial mass reservoir has been accreted onto the protostar plus disk system. Once the disk is formed, it occupies the innermost region of our numerical grid, while the infalling envelope---the remnant of the parent core---occupies the outer extent. The dynamics of both the disk and envelope are followed self-consistently on one global numerical grid space, which ensures correct mass infall rates onto the disk. This is an important prerequisite for studying gravitational instability and fragmentation in young protostellar disks at all epochs \citep[e.g.,][]{VB06,VB10,Machida10,Kratter10}.

We introduce a sink cell at $r_{\rm sc}=6.0$ AU and impose a free inflow inner boundary condition. In the early pre-stellar phase of evolution, we monitor the mass accretion rate through the sink cell and introduce a central point-mass object (representing the forming star) when the mass accretion rate reaches a peak value (see Fig.~\ref{fig9} for details). In the subsequent evolution, approximately 95\% of the accreted material lands directly onto the star, while the rest is maintained in the sink cell in order to keep its density equal to the mean density of the gas in the innermost 1--2 AU of the actual numerical grid.
The sink cell is otherwise dynamically inactive. It contributes only to the total gravitational potential, so that a smooth transition in column density is maintained from the numerical grid, into the sink cell, and to the protostellar surface.


We solve the mass and momentum transport equations, written in the thin-disk approximation as:
\begin{equation}
\label{cont}
{\partial \Sigma \over \partial t} +
{\bl \nabla}_{\rm p} \cdot \left(\Sigma {\bl v}_{\rm p} \right) = 0,
\end{equation}
\begin{equation}
\label{mom}
{\partial \over \partial t} \left( \Sigma {\bl v}_{\rm p} \right) + 
\left[
{\bl \nabla} \cdot \left(\Sigma {\bl v}_{\rm p} \otimes v_{\rm p}  \right)
\right]_{\rm p}
= - {\bl \nabla}_{\rm p} {\cal P} + \Sigma {\bl g}_{\rm p},
\end{equation}
where the subscript $p$ refers to the planar components $(r,\phi)$ in polar coordinates, $\Sigma$ is the mass surface density, ${\cal P}=\int^{Z}_{-Z} P dz$ is the vertically integrated form of the gas pressure $P$, $Z$ is the radially and azimuthally varying vertical scale height determined in each computational cell using an assumption of local hydrostatic equilibrium \citep{VB09}, $\bl{v}_{p}=v_r \hat{\bl r}+ v_\phi \hat{\bl \phi}$ is the velocity in the disk plane, $\bl{g}_{p}=g_r \hat{\bl r} +g_\phi \hat{\bl \phi}$ is the gravitational acceleration in the disk plane, and $\nabla_p=\hat{\bl r} \partial / \partial r + \hat{\bl \phi} \, r^{-1} \partial / \partial \phi $ is the gradient along the planar coordinates of the disk. The planar components of the divergence of the symmetric dyadic $\Sigma \bl{v}_p \otimes \bl{v}_p$ are found in \citet{VB10}. The thin-disk approximation is an excellent means of studying the evolution over many orbital periods and across a wide parameter space.

The gravitational acceleration $\bl{g}_p$ includes contributions from the central point-mass star (once formed), from material in the sink cell ($r<r_{\rm sc}$), and from the self-gravity of the circumstellar disk and envelope. The gravitational potential of the circumstellar disk and envelope is found  by solving the Poisson integral
\begin{eqnarray} 
\Phi(r,\phi) & = & - G \int_{r_{\rm sc}}^{r_{\rm out}} r^\prime dr^\prime \nonumber \\ 
 & & \times \int_0^{2\pi} \frac{\Sigma(r^\prime,\phi^\prime) d\phi^\prime}{\sqrt{{r^\prime}^2 + r^2 - 2 r r^\prime \cos(\phi^\prime - \phi) }}  \, ,
\end{eqnarray} 
where $r_{\rm out}$ is the radial position of the outer computational boundary, or equivalently, the initial radius of a pre-stellar core. This integral is calculated using a fast Fourier transform technique which applies the two-dimensional Fourier convolution theorem for polar coordinates with a logarithmically-spaced radial grid \citep[see][Sect.~2.8]{BT87}. 

\begin{figure}
  \centering
  \includegraphics[width=\columnwidth]{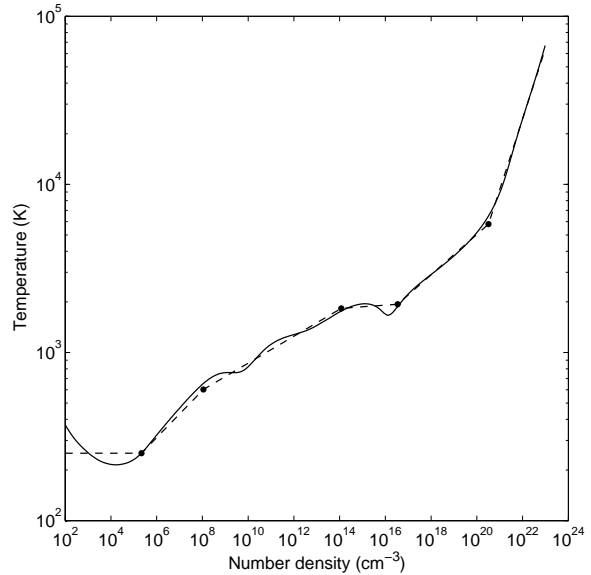}
  \caption{Temperature evolution of zero-metallicity gas derived from the one-zone calculations of \citet[][solid line]{Omukai05}, and the piecewise polytropic fit employed within the simulations discussed herein (dashed line); the filled circles mark the critical turning points of the fit specified in Table~\ref{table1}.}
  \label{fig1}
\end{figure}

Equations~(\ref{cont}) and (\ref{mom}) are closed with a barotropic equation of state, based on the 1D core collapse simulations of \citet{Omukai05} that included the detailed chemical and thermal processes of the collapsing gas. Figure~\ref{fig1} shows the gas temperature versus density relation from \citet{Omukai05} for zero metallicity (solid line) and our piecewise fit (dashed line), with the transition points denoted by filled circles. This piecewise polytropic form can be expressed as follows
\begin{equation}
P_k = c_{\rm s}^2 \rho^{\gamma_k} \prod_{i=1}^{k-1}\rho_{\rm{c},i}^{\gamma_i-\gamma_{i+1}},
\,\,\, \mathrm{for} \,\,\, \rho_{{\rm c},k-1} \le \rho < \rho_{{\rm c}, k} \, ,
\label{eosvol}
\end{equation}
where $c_{\rm s}=\sqrt{{\cal R} T/\mu}$ is the initial isothermal sound speed, $T$ is the initial gas temperature, $\cal R$ is the universal gas constant, and $\mu=2.27$ is the mean molecular weight 
of the primordial gas\footnote{True for gas with volume density higher than $10^{10}$~cm$^{-3}$. As a rule, our model disks are characterized by volume densities higher than this value, except perhaps in the very outer parts at $r>300-400$~AU.}. The value of the index $k$ distinguishes the six individual components of the
 piecewise form. We note that when $k=1$ the product term is unity, and the pressure reduces to $P_1=c_{\rm s}^2 \rho^{\gamma_1}$. As our simulations evolve the effective surface mass density of the gas, the corresponding form of the barotropic relation used in the code is
\begin{equation}
{\cal P}_k = c_{\rm s}^2 \Sigma^{\gamma_k} \prod_{i=1}^{k-1}\Sigma_{\rm{c},i}^{\gamma_i-\gamma_{i+1}},
\,\,\, \mathrm{for} \,\,\,  \Sigma_{{\rm c},k-1} \le \Sigma < \Sigma_{{\rm c}, k} \, ,
\label{eos}
\end{equation}
where the transition surface and volume mass density are related to one another through the instantaneous local scale height $Z$ at each point in the disk via $\rho_{{\rm c},i}=\Sigma_{{\rm c},i}/2Z$. The transition points $k$, the associated mass and number volume densities at which these transitions occur, $\rho_{\rm{c},i}$ and $n_{{\rm c},i}$, and the value of the various associated polytrope indices are given in Table~\ref{table1}. Note that $\Sigma_{{\rm c},0}$ and $\Sigma_{{\rm c},6}$ in equation~(\ref{eos}) are formally equal to zero and infinity, respectively.

\begin{table}
\begin{center}
\caption{Parameters of the Barotropic Relation}
\label{table1}
\renewcommand{\arraystretch}{1.5}
\begin{tabular*}{\columnwidth}{ @{\extracolsep{\fill}} c c c c }
\hline \hline
$k$ & $\gamma_i$ & $\rho_{\rm{c},i}$ & $n_{\rm{c},i}$ \\
\hspace{1cm} & & (g~cm$^{-3}$)   & (cm$^{-3}$) \\ [0.5ex]
\hline \\ [-2.0ex]
1 & 1.00 & $8.20{\times}10^{-19}$ & $2.16{\times}10^{5}$ \\
2 & 1.14 & $4.19{\times}10^{-16}$ & $1.10{\times}10^{8}$ \\
3 & 1.08 & $4.50{\times}10^{-10}$ & $1.18{\times}10^{14}$ \\
4 & 1.01 & $1.32{\times}10^{-7}$ & $3.47{\times}10^{16}$ \\
5 & 1.12 & $1.23{\times}10^{-3}$ & $3.24{\times}10^{20}$ \\
6 & 1.42 & --- & --- \\ [1.0ex]
\hline
\end{tabular*}
\end{center}
\end{table}

The initial gas surface density $\Sigma$ and angular velocity $\Omega$ profiles for the primordial cores are similar to those that have been considered in the context of present-day star formation \citep{VB06,VB10}
\begin{equation}
\label{ic1}
\Sigma = {r_0 \Sigma_0 \over \sqrt{r^2+r_0^2}}\:,
\end{equation}
\begin{equation}
\label{ic2}
\Omega = 2 \Omega_0 \left( {r_0\over r}\right)^2 \left[\sqrt{1+\left({r\over r_0}\right)^2} -1\right].
\end{equation}
The radial profile of $\Sigma$ is an integrated form of a Bonnor-Ebert sphere \citep{Dapp09}, while that of $\Omega$ is the expected differential rotation profile to accompany (\ref{ic1}) for a core contracting from near-uniform initial conditions \citep{Basu97}. The parameters $\Omega_0$, $\Sigma_0$, and $r_0$, are the central angular velocity, central gas surface density, and the radius of a central near-constant-density plateau, respectively. The latter is proportional to the Jeans length and is defined as
\begin{equation}
\label{rzero}
r_0 = {\sqrt{A} c_{\rm s}^2 \over \pi G \Sigma_0}. 
\end{equation}
The parameter $A$ determines the level of gravitational domination in the initial state. It is set to 1.5 for all models considered herein, and the initial gas temperature is set to 250~K (unless otherwise stated). We note that we have previously shown that the qualitative features of the protostellar accretion and disk evolution are weakly sensitive to the specific profiles adopted for the initial prestellar core \citep{VB09,Vorobyov12}.

To set up a model prestellar core using equations~(\ref{ic1}) and (\ref{ic2}), the values of $\Sigma_0$ and $\Omega_0$ need to be specified. The former is found from equation~(\ref{rzero}) by assuming a constant ratio of the core radius $r_{\rm out}$ to the radius of the central plateau $r_0$: $r_{\rm out}/r_0=6$. All model cores therefore possess a similarly truncated form. The parameter $r_0$ is chosen so as to form cores with mass on the order of $200~\mbox{M}_\odot$. This value is typical for primordial collapsing starless cores, as found in the numerical hydrodynamics simulations of \citet{Yoshida06}.

The central angular velocity $\Omega_0$ of our model cores are characterized by the dimensionless parameter $\eta \equiv \Omega_0^2 r_0^2/c_{\rm s}^2$ \citep{Basu97}, which is related to the ratio $\beta=E_{\rm rot}/|E_{\rm grav}|$ of rotational to gravitational energy by $\beta\,{\approx}\,0.9\eta$. The value of the central angular velocity $\Omega_0$ is then calculated by choosing a value of $\beta$ appropriate for primordial cores with spin parameter $\alpha=\sqrt{\beta}\,{\approx}\,0.05$ \citep{OShea07}. Table~\ref{table2} shows a list of parameters for the models presented in this study.

\begin{table}
\begin{center}
\caption{Model Parameters}
\label{table2}
\renewcommand{\arraystretch}{1.5}
\begin{tabular*}{\columnwidth}{ @{\extracolsep{\fill}} *{7}c }
\hline \hline
Model & $\Sigma_0$ & $\Omega_0$ & $r_0$ & $M_{\rm c}$ & $\beta$ & $T$ \\
 & g~cm$^{-2}$  & (km~s$^{-1}$~pc$^{-1}$) & (pc) & ($\mbox{M}_\odot$) & ($10^{-3}$) & (K) \\ [0.5ex]
\hline \\ [-2.0ex]
\textit{ref.} & 0.26 & 0.75 & 0.4 & 176 & 2.76 & 250 \\
\textit{1} & 0.26 & 0.375 & 0.4 & 176 & 0.69 & 250 \\
\textit{2} & 0.5 & 0.75 & 0.29 & 179 & 1.97  & 350 \\ [1.0ex]
\hline
\end{tabular*}
\end{center}
\end{table}

All models are run on a polar coordinate ($r,\phi$) grid with $512 \times 512$ spatial zones. The inner and outer boundary conditions are set to allow for free outflow from the computational domain. The radial points are logarithmically spaced, allowing for better numerical resolution of the inner grid, where the disk forms and evolves. In the reference model, the innermost cell outside the central sink has a radius of 0.11~AU and the radial and azimuthal resolution are about 1.9~AU at a radius of 100 AU. This resolution is sufficient to fulfil the Truelove criterion, which states that the local Jeans length must be resolved by at least four numerical cells \citep{Truelove99}.

Indeed, the Jeans length of a thin self-gravitating disk can be written as
\begin{equation}
\label{RJeans}
R_{\rm J} = { c_{\rm s}^2 \over G \Sigma_0 }.
\end{equation}
For a surface density of $\Sigma_0\,{\approx}\,500-1000~\mbox{g~cm}^{-2}$ and temperature $T\,{\approx}\,1.0-1.5{\times}10^3~\mbox{K}$, typical for our disks at $r\,{\approx}\,100~\mbox{AU}$ (see Figs.~\ref{fig1} and \ref{fig7}), the corresponding Jeans length varies between $R_{\rm J}{\approx}35-90~\mbox{AU}$ and is resolved by roughly 17--45 grid zones in each direction $(r,\phi)$.

Finally, we want to emphasize that we do not resort to the use of sink particles of any sort except for the central sink cell. The fragments studied in this paper are fully self-gravitating protostellar embryos supported against gravity by pressure and rotation. 

\begin{figure}
  \resizebox{\hsize}{!}{\includegraphics{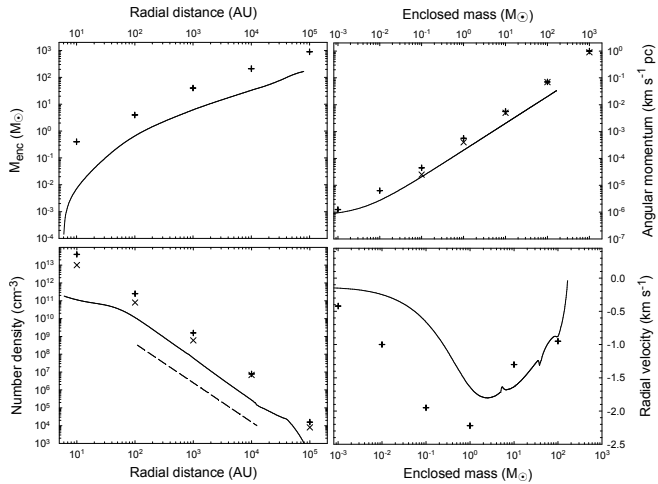}}
  \caption{Comparison of the core properties in the reference model with those found using numerical hydrodynamics simulations of collapsing primordial mini-halos. The solid lines present the radial profiles of the enclosed mass (top-left panel) and gas volume density (bottom-left panel), and also the specific angular momentum and radial velocity vs. enclosed mass (top-right and bottom-right panels, respectively). The plus signs and crosses are the corresponding data taken from \citet{Clark2011} and \citet{Yoshida06}, respectively. The dashed line shows the $r^{-2.2}$ profile, typical for the volume density in collapsing mini-halos. }
  \label{fig2}
\end{figure}

\begin{figure}
  \resizebox{\hsize}{!}{\includegraphics{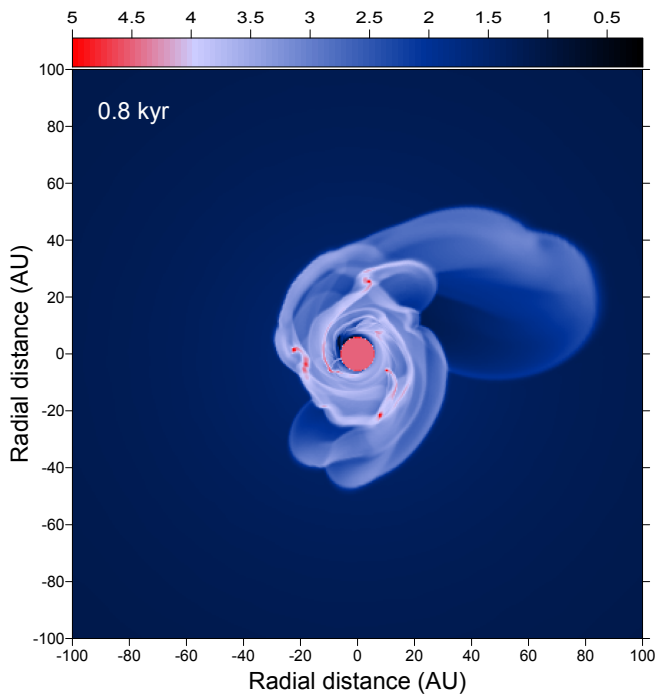}}
  \caption{Gas surface density map in the reference model showing the disk when its age is just 800~yr. The scale bar is in log~g~cm$^{-2}$. Multiple fragments are already evident in the disk at this early time.}
  \label{fig3}
\end{figure}

\begin{figure*}
  \centering
  \includegraphics[width=16cm]{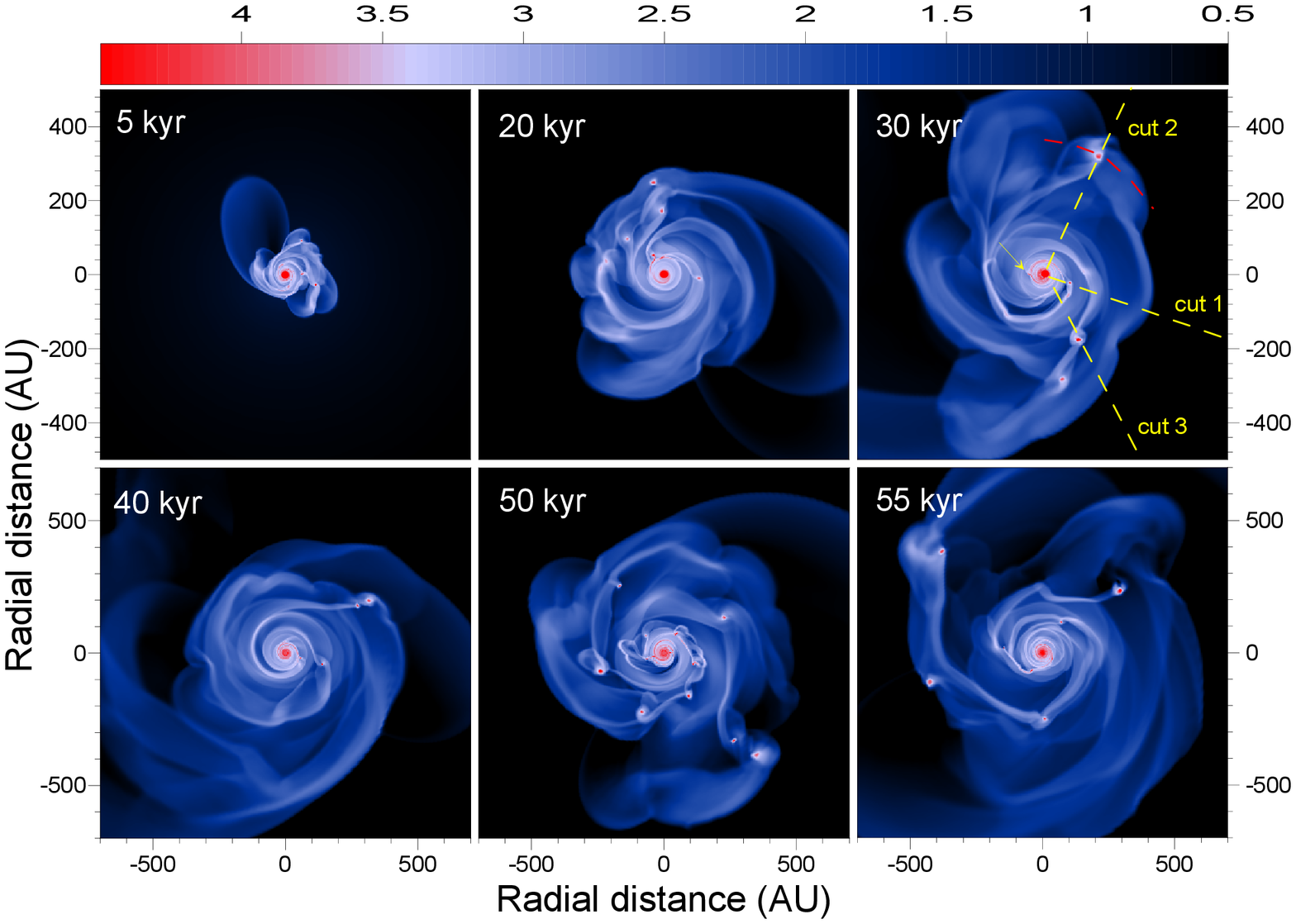}
  \caption{Gas surface density maps of the circumstellar disk in the reference model. The time elapsed since the formation of the central star is indicated in each panel. The scale bar is logarithmic in $\mbox{g~cm}^{-2}$. The yellow dashed lines indicate radial cuts passing through several fragments that are used to calculate the radial gas surface density profiles in Fig.~\ref{fig7}. The red dashed curve illustrates an azimuthal cut used to calculate the gravitational torques acting on a fragment in Fig.~\ref{fig7}, and the arrow points to fragment F4 in Fig.~\ref{fig8}.}
  \label{fig4}
\end{figure*}

\section{Results}
\label{results}
In this section we consider the time evolution of our reference model (see Table~\ref{table2}), starting from the prestellar phase and ending once the mass of the central star reaches $45~\mbox{M}_\odot$. Beyond this mass, the effect of stellar irradiation may strongly affect our results \citep{Hosokawa2011}.

\subsection{Cloud core at the onset of the formation of the central protostar} 
We start by comparing the properties of our collapsing core in the reference model just prior to the formation of the central star to those derived using three-dimensional numerical hydrodynamics simulations of collapsing primordial mini-halos. The solid lines in Figure~\ref{fig2} show the enclosed mass $M_{\rm enc}$ vs. radial distance $r$ (top-left panel), the gas volume density $n_{\rm g}$ vs. $r$ (bottom-left panel), the specific angular momentum $L$ vs. $M_{\rm enc}$ (top-right panel), and the radial velocity $v_{\rm r}$ vs. $M_{\rm enc}$ (bottom-right panel). The volume density was retrieved from the gas surface density $\Sigma$ and the vertical scale height $Z$ using the relation $n_{\rm g}=\Sigma/(2Z\mu m_{\rm H})$. We compare our model profiles with those of \citet{Clark2011} (plus signs) and and \citet{Yoshida06} (crosses). The form of the core in the reference model is similar to those derived using 3D simulations. For instance, the density profile follows closely the usual $r^{-2.2}$ form (shown by the dashed line to guide the eye). A steeper falloff in the core outer regions is caused by a finite core boundary \citep{VB05b}. However, notable differences are seen in the actual values of $n_{\rm g}$, $M_{\rm enc}$, $v_{\rm r}$, and $L$. For instance, the gas density in our model is about a factor of several lower than that of \citet{Yoshida06} and about an order of magnitude lower than that of \citet{Clark2011}. This is primarily caused by the fact that these authors considered more massive cores than we did in our study. This is evident from the top-left panel showing the enclosed mass as a function of radius. The specific angular momentum in our model is a factor of several smaller than that of \citet{Clark2011} but is approaching that of \citet{Yoshida06} in the inner region. We could have reconciled the numerically derived profiles by considering cores with a higher initial positive density perturbation $A$ and/or higher angular momentum. This, however, would not alter our main results because such cores would produce more massive and extended disks, whose properties would favor gravitational fragmentation even more \citep[e.g.][]{VB06,VB10}.

\subsection{Formation and evolution of a primordial disk}

The disk in the reference model starts to form at about $t=1.0$~kyr after the formation of the central protostar. This delay between the formation of the protostar and the disk could have been even shorter had the sink cell radius been smaller than 6~AU. The first episodes of fragmentation occurred about 300--400~yr after the formation of the disk. Figure~\ref{fig3} shows the gas surface density (in log~g~cm$^{-2}$) in the inner $200\times200$ box when the disk age is only 800~yr. The red circle in the coordinate center represents the sink cell. Several fragments have already formed by this time. This fragmentation timescale is longer than that found by \citet{Clark2011} and \citet{Greif2011}, who used sink cells with a smaller radius of 1.5~AU, but is comparable to the fragmentation timescale found by \citet{Smith2011}, who used larger sink cells (20~AU). We note that our core is initially of lower density and angular momentum than in those studies, which may act to increase the disk fragmentation timescale in our simulations.

Figure~\ref{fig4} presents a series of images of the gas surface density (logarithmic in g~cm$^{-2}$) for the inner thousand AU\footnote{The top row has a spatial scale of $1000{\times}1000$ AU, while the bottom row has a scale of $1400{\times}1400$ AU.}. The time elapsed since the formation of the central star is indicated in each panel. We note that the whole computational domain extends to roughly 80,000 AU (exactly, 0.4~pc) but that we focus on the innermost regions where the disk forms around the central protostar.
The number of fragments present in the disk varies with time, indicating that they are not long lived, either migrating onto the star or being dispersed. In both cases, gravitational torques from spiral arms and/or other fragments are responsible. However, new fragmentation episodes take place because the disk is constantly supplied with material from the parent core. Most of the fragments form in the intermediate and outer disk regions, in agreement with numerical models of disk evolution in the present-day universe \citep[e.g.,][]{Stamatellos08,Clarke09}.

\begin{figure}
  \resizebox{\hsize}{!}{\includegraphics{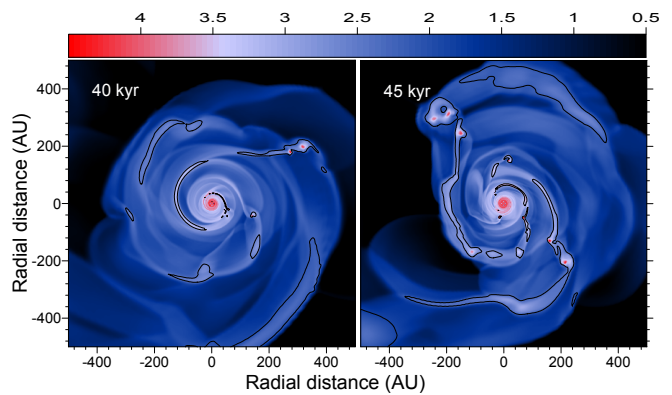}}
  \caption{Gas surface density maps for the reference model shown at $t=40$~kyr (left-hand panel) and $t=45$~kyr (right-hand panel) after the formation of the central star. The contour lines delineate regions in which the Toomre $Q$-parameter is lower than unity. The scale bar is in log~g~cm$^{-2}$.}
  \label{fig5}
\end{figure}

\begin{figure}
  \centering
  \includegraphics[width=\columnwidth]{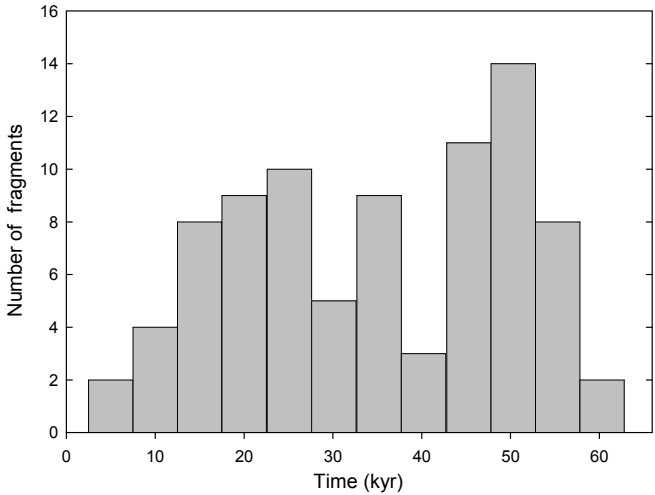}
  \caption{Number of fragments in the disk at a given time, as seen in the reference model.}
  \label{fig6}
\end{figure}

The recurrent character of disk fragmentation is evident in Figure~\ref{fig6} in which we plot the number of fragments $N_{\rm f}$ present in the disk with time. $N_{\rm f}$ increases steadily to 10 by $t=25~\mbox{kyr}$, and then suddenly decreases by $t=30~\mbox{Myr}$ to $N_{\rm f}=5$, suggesting that half of the fragments have migrated or been dispersed in just $5~\mbox{kyr}$. This is followed by a period of intense fragmentation followed by another rapid depletion during the subsequent $10~\mbox{kyr}$. During the most vigourous episode of disk fragmentation the number of fragments in the disk peaks at $N_{\rm f}=14$ at $t=50~\mbox{kyr}$. This fragmentation burst is followed by a deep minimum when the number of fragments drops to  just $N_{\rm f}=2$. It is important to note that $N_{\rm f}$ never drops to zero, suggesting that some of the fragments may not migrate inward but instead stay at quasi-stable wide orbits or even migrate outward as also found in recent studies by \citet{Clark2011} and \citet{Greif2011}.

The consecutive increases and declines in the number of fragments suggests that the disk approaches a limit cycle during which periods of vigourous disk fragmentation are followed by periods of relatively weak (to no) fragmentation. This limit cycle behaviour is possible if the typical migration time scale of the fragments is shorter than the characteristic time required for disk loading of material infalling from the collapsing parent core. The disk therefore loses mass (via inward migration of the fragments onto the star) faster than it can be replenished via accretion from the parent core. Following significant mass loss, the disk stays relatively dormant until sufficient mass is accumulated to trigger another burst of fragmentation.

For a mass infall rate onto the disk of $\dot{M}_{\rm d}\approx10^{-3}~\mbox{M}_\odot~\mbox{yr}^{-1}$ and disk mass $M_{\rm d}\,\approx\,10~\mbox{M}_\odot$ (see Fig.~\ref{fig9}), the characteristic timescale for disk mass loading is
\begin{equation}
t_{\rm l}={M_{\rm d} \over \dot{M}_{\rm d}} \approx 10~\mathrm{kyr}.
\end{equation}
The migration timescale $t_{\rm m}$ estimated for fragments in the reference model lies in the range of $10^3-7\times10^4~\mbox{yr}$, with most fragments having migration times of just a few thousand years (see Table~\ref{table3} and equation~[\ref{migtime}]). Hence, the mean migration time is indeed shorter than the characteristic mass loading time. It is worth noting that $t_{\rm l}$ is in approximate agreement with the temporal variability of $5-15~\mbox{kyr}$ over which peaks in the distribution of fragment numbers occur in Figure~\ref{fig6}.

To illustrate the effect of gravitational fragmentation, regions in which the Toomre $Q$-parameter fall below unity are plotted in Figure~\ref{fig5} (traced in black). The $Q$-parameter is defined as $c_{\rm s} \Omega/\pi G \Sigma$. Evidently, all fragments are characterized by $Q<1$.  In addition, some parts of the spiral arms are also characterized by $Q<1$ but do not yet show signs of fragmentation. Indeed, the number of fragments at $t=40$~kyr (right-hand panel)  is greater than at $t=45$~kyr (left-hand panel), indicating that new episodes of disk fragmentation took place during the intervening period. We note that the $Q$-parameter in  the innermost 50~AU is greater than unity, which explains the stability of this region to gravitational fragmentation.

\begin{figure}
  \centering
  \includegraphics[width=\columnwidth]{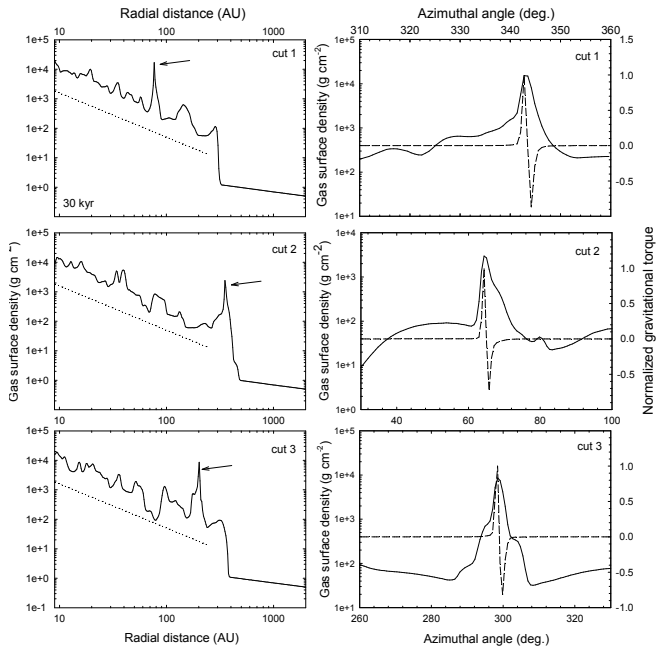}
  \caption{{\bf Left column:} Radial gas surface density profiles calculated along the radial cuts passing through several individual fragments shown in Fig.~\ref{fig4} by yellow dashed lines. The number of the corresponding cut is indicated in each panel. Arrows point to the radial position of the fragments. The dotted lines show an $r^{-1.5}$ radial profile for comparison. {\bf Right column:} The azimuthal profiles of the gas surface density (solid lines) and normalized gravitational torque (dashed lines) calculated along the azimuthal cuts passing through the same fragments. One of these cuts is shown schematically in Fig.~\ref{fig4} by the red dashed curve.}
  \label{fig7}
\end{figure}

The left-hand column of Figure~\ref{fig7} presents the gas surface density distributions calculated along the radial cuts shown schematically by yellow dashed lines in the upper-left panel of Figure~\ref{fig4} ($t=30$~kyr). Each cut has an angular width of $15^\circ$, and is centered on the peak density of the fragment and and its surrounding minidisk. The gas surface density is averaged over the grid zones that are overlayed by the cut and have the same radial distance. The top, middle and bottom rows correspond to cuts 1, 2, and 3, respectively. The positions of the fragments are indicated by arrows. The characteristic $r^{-1.5}$ slope is shown as a dotted line for convenience. Evidently, the radial profile of the gas surface density scales as $\Sigma\propto r^{-1.5}$, which is typical for self-gravitating disks in present-day star formation \citep[e.g.,][]{VB2007,VB09}.
The typical gas volume density averaged over the azimuthal angle is a few~$\times 10^{14}$~cm$^{-3}$ at $r=10$~AU and it drops to about a few~$\times 10^{11}$~cm$^{-3}$ beyond 100~AU. These values are in reasonable agreement with three-dimensional numerical hydrodynamics simulations of disk formation around primordial
protostars \citep{Clark2011}.

The right-hand side column in Figure~\ref{fig7} presents the gas surface density distributions (solid lines) and normalized gravitational torques (dashed lines) calculated along the {\it azimuthal} cuts passing through each of the fragments. One such cut is shown by the red dashed curve in the upper-right panel of Figure~\ref{fig4} for illustration. The width of the cuts is 10 grid zones, which, depending on the radial position, translates into a radial width of 15--55~AU. The gravitational torque is calculated as $\tau = - m(r,\phi)\, \partial \Phi / \partial \phi$, where $m(r,\phi)$ is the gas mass in a cell with polar coordinates $(r,\phi)$. The gas surface density and integrated torque are averaged over the grid zones that are overlayed by the cut and have the same radial distance.

Figure~\ref{fig7} shows that the gas surface density in fragments is higher by about two orders of magnitude as compared to local disk values. The behavior of the normalized gravitational torque is interesting in that it exhibits a steep rise to a maximum {\it positive} value  on the leading part of the fragment (greater azimuthal angles) and a deep drop to a minimum {\it negative} value on the trailing part (smaller azimuthal angles). The absolute value of $\tau$ is several orders of magnitude lower everywhere else. This specific form of $\tau$ in profile is caused by tidal forces acting on the fragment from the rest of the disk and, in particular, from the  spiral arms within which the fragments are usually nested. The trailing part of the arm (with respect to the fragment) exerts a negative gravitational torque and pulls the fragment back, while the leading part of the arm exerts a positive torque and pulls the fragment forward in the direction of rotation. The resulting tidal force tends to shear apart the fragment, but the fragment's own self-gravity prevents it from dispersing.

To calculate the properties of the fragments, we designed an algorithm that located the fragments in the disk and calculated their mass $M_{\rm f}$, maximum surface density $\Sigma_{\rm max}$, physical and Hill radii, $R_{\rm f}$ and $R_{\rm H}$, mass within the Hill radius $M_{\rm H}$, radial distance of the fragment $r_{\rm f}$, and the integrated gravitational torque acting on the fragments $\tau_{\rm f}$. In particular, the Hill radius is
\begin{equation}
R_{\rm H } = r \left( {M_{\rm f} \over 3(M_{\rm f} + M_\ast)}  \right)^{1/3},
\end{equation}
where $M_\ast$ and $M_{\rm f}$ are the masses of the star and fragment, respectively. The radius of the fragment $R_{\rm f}$ is calculated from the known area occupied by the fragment, assuming a circular shape. Additional details of the tracking algorithm are described in the Appendix.

\begin{figure*}
  \centering
  \includegraphics[width=16cm]{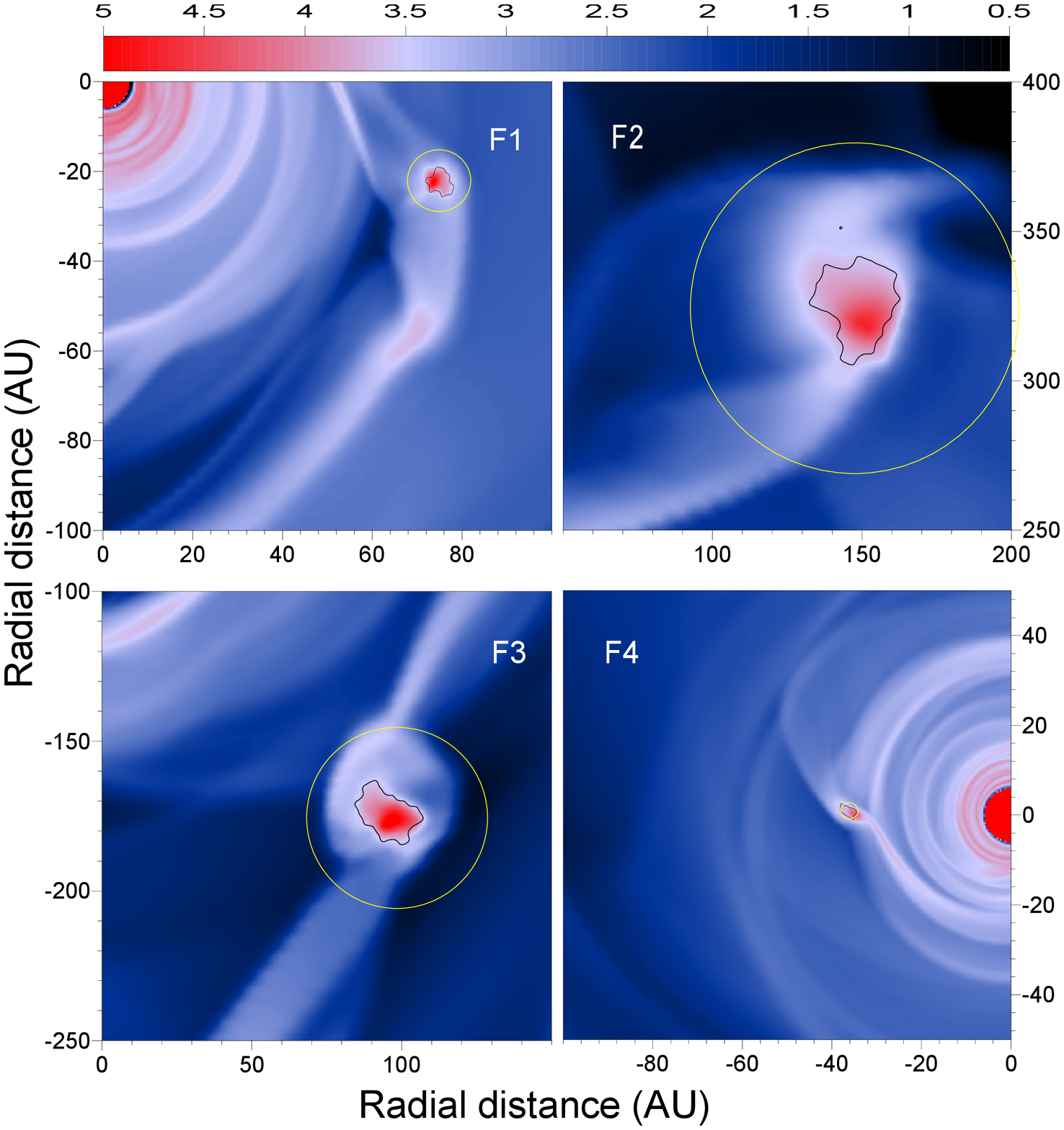}
  \caption{Zoom-in images of four typical fragments present in the disk at $t=30$~kyr in Fig.~\ref{fig4}. Fragments F1--F3 are named after the radial cuts that pass through the corresponding fragments, while fragment F4 is indicated by the arrow. The black contour lines delineate the fragments as determined by our fragment-tracking algorithm, while the yellow line outlines the Hill radius. 
The scale bar is logarithmic in g~cm$^{-2}$.}
  \label{fig8}
\end{figure*}

Figure~\ref{fig8} presents zoomed-in gas surface density maps of four typical fragments. Fragments F1--F3 are named for the radial cuts denoted by the yellow dashed lines passing through the corresponding fragments in Figure~\ref{fig4}. Fragment~F4 is marked in Figure~\ref{fig4} by the arrow. Fragment positions, as determined by our fragment-tracking algorithm, are outlined in black, and their properties listed in Table~\ref{table3}.


All of the individual fragment radii $R_{\rm f}$ are smaller than their respective Hill radii $R_{\rm H}$ (see Fig.~\ref{fig8}), indicating that the fragments are gravitationally bound objects. Fragment masses $M_{\rm f}$ lie in the range $0.02-0.38~\mbox{M}_\odot$. As the mass located within the Hill radius $M_{\rm H}$ is always greater than $M_{\rm f}$, the excess material can be interpreted as a minidisk. The presence of minidisks is most evident around fragments F1--F3 in Figure~\ref{fig8}. Fragment F4 is interesting in that its radius is nearly equal to the Hill radius, which is a manifestation of tidal stripping; it is also the least massive of the four fragments. In general, the closer the fragment is located to the central protostar, the smaller its radius and mass.

The orbital dynamics of the fragments depends on the sign of the integrated gravitational torque acting on them. As indicated from the data in Table~\ref{table3}, all of the fragments in Figure~\ref{fig8} migrate inward.  We note that occasional outward migration of the fragments as a result of N-body-like gravitational scattering was also found in our models. The migration time scale can be estimated as follows. A (small) change in the angular momentum of a fragment (mass $M_{\rm f}$) on a Keplerian orbit, due to a (small) change in the orbital distance of the fragment $dr_{\rm f}$ can be written as $dL=\frac{1}{2}M_{\rm f} v_{\rm f} dr_{\rm f}$, where $v_{\rm f}=(GM_\ast/r_{\rm f})^{1/2}$. The migration velocity of the fragment is then
\begin{equation}
v_{\rm m}={dr_{\rm f}\over dt} = {2 {dL \over dt} \over M_{\rm f} v_{\rm f}}.
\label{migtime}
\end{equation}
Noticing that $dL/dt={\cal T}$, where the latter quantity is calculated as the sum of all individual gravitational torques $\tau$ acting on the fragment, the characteristic migration time is
\begin{equation}
t_{\rm m}= {r_{\rm f} \over v_{\rm m}}= {L \over 2 {{\cal T}}}.
\label{migrate}
\end{equation}
The estimated migration time scale from each fragment is listed in Table~\ref{table3}. Fragment F4 and F2 have the shortest/longest migration time scales, of $10^3~\mbox{yr}$ and $7\times10^4~\mbox{yr}$, respectively. We note that these values are {\it instantaneous} migration time scales, and that their migration patterns may change as more envelope material accretes on to the disk. Nevertheless, these migration time scales indicate that most of the fragments that are seen in the disk at $t=30~\mbox{kyr}$ will eventually have migrated onto the forming protostar by the end of our simulations, at $t=55~\mbox{kyr}$. The number of fragments in Figure~\ref{fig4} does not steadily decline in time because subsequent generations of disk fragmentation continue to be fuelled by mass loading from the infalling envelope (e.g., at $t=50$~kyr).

Ultimately, fragments that survive tidal stripping and/or shearing will pass through the inner boundary of the simulation, after which the evolution may branch into one of two possibilities. If the fragment survives tidal stripping and shearing, it may settle into a close orbit and form a low-mass companion close to the central star. Alternatively, the fragment may merge with or be tidally destroyed by the central star, releasing its gravitational energy in the form of a strong luminosity outburst. The latter possibility has been suggested by the recent numerical hydrodynamics simulation of \citet{Greif2012}. We consider this phenomenon in more detail below.

\begin{table*}
\begin{center}
\caption{Fragment Characteristics}
\label{table3}
\renewcommand{\arraystretch}{1.5}
\begin{tabular}{ @{\extracolsep{\fill}} *{10}c }
\hline \hline
Fragment & $M_{\rm f}$ & $M_{\rm H}$ & $R_{\rm f}$ & $R_{\rm H}$ & $\Sigma_{\rm max}$ & 
$T_{\rm max}$ &  $r_{\rm f}$ & $\tau_{\rm f}$ & $t_{\rm m}$ \\
& $(\mbox{M}_\odot)$  &  $(\mbox{M}_\odot)$ & (AU) &   (AU)& (g~cm$^{-2}$) & (K) & (AU) & (g~cm$^2$~s$^{-2}$) & (yr) \\ [0.5ex]
\hline \\ [-2.0ex]
F1 & 0.08 & 0.1 & 3.5 & 7.3 & $1.3\times 10^5$ & 1890  &  77 & -13.2 & $4.1\times 10^3$ \\
F2 & 0.38 &  0.54 & 23.0 & 56.1 & $1.7\times 10^4$ & 1800 &  352 & -3.0 & $7\times 10^4$ \\
F3 & 0.35 &  0.46 & 12.0 & 31.1 & $6.2\times 10^4$ & 1860 & 202 & -11.4 & $1.3\times 10^4$ \\
F4 & 0.02 &  0.02 & 1.83 & 2.0 & $3.0\times 10^4$ & 1840 & 35 & -2.9 & $10^3$ \\ [1.0ex]
\hline
\end{tabular}
\tablecomments{Torque $\tau_{\rm f}$ is in units of $8.6\times 10^{40}$}
\end{center}
\end{table*}
 
\subsection{Accretion and luminosity bursts}
\begin{figure}
  \centering
  \includegraphics[width=10 cm]{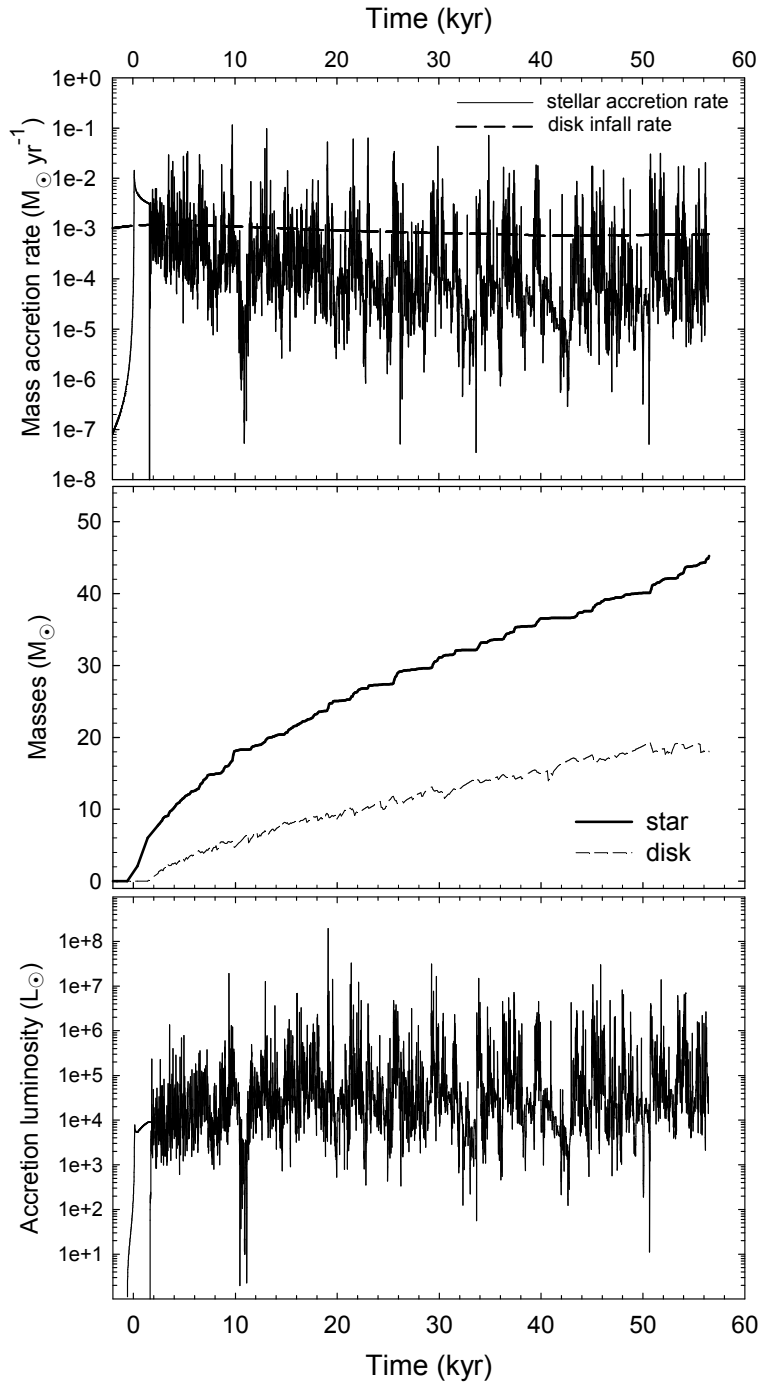}
  \caption{\textbf{Top:} Mass accretion rates, from the disk onto the central protostar (solid), and from the remnant cloud core onto the disk at 3000 AU (dashed). \textbf{Middle:} Temporal evolution of the masses of the protostar and its disk. Punctuated increases in the protostar mass, associated with burst events, correspond to the momentary decreases in disk mass. \textbf{Bottom:} Accretion luminosity associated with the mass accretion rate onto the protostar.}
  \label{fig9}
\end{figure}

Temporal evolution of the mass accretion rate in the reference model is shown in the top panel of Figure \ref{fig9}, from the disk onto the protostar (solid line), together with the smoothly varying accretion rate of material from the envelope onto the disk at 3000 AU (dashed line). The formation of the protostar is marked by the sharp peak in the mass accretion rate; we fix this point as $t=0$ and count time forward from here. The protostar then accretes material from the immediate surrounding envelope at the rate of a few times $10^{-3}~\mbox{M}_{\odot}~\mbox{yr}^{-1}$, growing to $6.0~\mbox{M}_{\odot}$ within approximately $2~\mbox{kyr}$. At this point the accretion is temporarily halted by the formation of a quasi-Keplerian disk.

Accretion quickly resumes as the disk grows in mass and the influence of gravitational torques acting within the disk continues to redistribute matter and angular momentum. We calculate the mean accretion rate in 1,000 year intervals following the formation of the quasi-Keplerian disk, finding $\dot{M}$ to decline steadily from approximately $10^{-3}$ to $10^{-4}~\mbox{M}_{\odot}~\mbox{yr}^{-1}$ over the course of the simulation. However, this smoothed background of accretion is punctuated by significant burst events during which clumps of typically ${\sim}0.03~\mbox{M}_{\odot}$ in size are accreted on time scales of less than 100 yr (e.g., fragment F4 in Figure~\ref{fig8}). A select few individual burst events involve the accretion of clumps of $\sim 1.0~\mbox{M}_\odot$. This results in effective mass accretion rates of ${\sim}10^{-1}~\mbox{M}_{\odot}~\mbox{yr}^{-1}$, with the most extreme rates as high as $1.0~\mbox{M}_{\odot}~\mbox{yr}^{-1}$. Such large fluctuations in the accretion rate are the characteristic signature of gravitationally-induced fragmentation within the disk, and subsequent accretion of the fragments onto the protostar.

The cumulative effect of these two disparate modes of accretion are summarized by the mass growth curves in the middle panel of Figure \ref{fig9}. The protostar mass (solid line) grows rapidly over ${\sim}5~\mbox{kyr}$ during an initial phase of smooth accretion, continuing to increase steadily thereafter. However, the burst mode of accretion is also evident through abrupt increases in mass, of a few $\mbox{M}_{\odot}$ at a time. These increases are typically followed by plateaus during which the mass of the protostar changes very little while the disk equilibrates back into a quasi-Keplerian state. Each burst event is also mirrored by a corresponding decrease in the total disk mass (dashed line). Overall however, the disk mass continues to increase due to the accretion of material from the remnant parent cloud core.

An object's luminosity during the early stages of protostellar evolution can be effectively characterized by the accretion luminosity alone. This is estimated by the dissipation of kinetic energy from infalling disk material landing on the protostellar surface:
\begin{equation}
\label{eqn:accretionluminosity}
L_{\rm acc}
=
\frac{GM_{*}\dot{M}}{2 R_{*}},
\end{equation}
where $M_{*}$ is the mass of the protostar, $\dot{M}$ is the accretion rate from the disk, and $R_{*}$ is the protostellar radius. To determine $R_{*}$ in the absence of a detailed model for the stellar interior, we adopt the evolutionary model of \citet{OP2003}, and following \citet{Smith2011}, employ simple power-law approximations for the protostar radius during each phase of its evolution.

Following the initial collapse of the cloud core, \citet{SPS1986} showed that the protostellar radius grows according to a mixed power-law of the form $R_{*}{\propto}M_{*}^{0.27}\dot{M}_{-3}^{0.41}$; for notational convenience we use $\dot{M}_{-3}$ to denote the ratio of the actual mass accretion rate $\dot{M}$ to the fiducial value of $10^{-3}~\mbox{M}_{\odot}~\mbox{yr}^{-1}$ \citep{OP2003}. However, this growth is exacerbated once the internal temperature of the protostar rises sufficiently to drive a wave of luminosity outward from the core. The result is a sudden and rapid expansion of the stellar surface. Once this wave reaches the surface of the star itself, the interior is able to relax, and the protostar begins Kelvin-Helmholtz contraction toward the main-sequence. The following power-law relations therefore approximate the evolution of $R_{*}$ through these distinct phases, as well as the transitions between them \citep{Smith2011}:
\begin{equation}
\label{eqn:radiusrelations}
R_{*} = \left\{
\begin{array}{ll}
26 M_{*}^{0.27} \dot{M}_{-3}^{0.41}, & M_{*}~{\leq}~M_1 \\
A_1 M_{*}^3, & M_1~{\leq}~M_{*}~{\leq}~M_2 \\
A_2 M_{*}^{-2}, & M_2~{\leq}~M_{*}~\&~R_{*}<R_{\rm ms}
\end{array} \right..
\end{equation}
The constants $A_1$ and $A_2$ are matching conditions that ensure the functional form of $R_{*}$ remains smoothly varying during the transitions.

The mass parameter $M_1$ marks the transition between the adiabatic phase of growth and the arrival of the luminosity wave at the protostellar surface; $M_2$, the transition between the luminosity wave driven expansion and subsequent Kelvin-Helmholtz contraction. $M_1$ and $M_2$ are fixed by the instantaneous mass accretion rate as the protostar transitions between phases, and are defined as
\begin{equation}
\label{eqn:radiitransitionpts}
\begin{array}{c}
M_1 = 5{\dot{M}_{-3}}^{0.27}, \\
M_2 = 7{\dot{M}_{-3}}^{0.27}.
\end{array}
\end{equation}

Although this model was originally developed under the assumption of a constant mass accretion rate, evolution of the interior can be assumed to occur roughly adiabatically due to the long cooling time of the protostellar interior. As a result, significant rapid variability in $R_{*}$ is not expected while the Kelvin-Helmholtz time scale is much longer than the accretion time scale.

The bolometric accretion luminosity $L_{\rm acc}$ for our reference model is presented in the bottom panel of Figure~\ref{fig9}. $L_{\rm acc}$ reaches a value on the order of $10^{4}~\mbox{L}_{\odot}$ almost immediately at the time the protostar is formed, and continuing to vary slowly about this level over the course of the simulation, typical for protostars of primordial composition \citep[e.g.,][]{Smith2012}. After the disk is formed at $t=2~\mbox{kyr}$, the accretion luminosity demonstrates a highly variable pattern of behaviour with high-intensity bursts (up to $10^7~\mbox{L}_{\odot}$) superimposed on the lower baseline value ($\sim 10^4~\mbox{L}_{\odot}$). 
This highly variable luminosity may have important
consequences for the disk evolution, but more accurate numerical
simulations taking into account the heating/cooling balance are needed to
assess this effect. 

Spiral arms are formed when the destabilizing effects of self-gravity become comparable to, or dominate over, the stabilizing effects of pressure and shear. If the Toomre $Q$-parameter within the arm drops below unity, sections of the arm may collapse to form bound fragments. The accretion bursts occur when these fragments are driven onto the protostar by the gravitational torques of the spiral arms and/or other fragments. The temporal evolution of $Q$ and the integrated torque $\cal T$ can be used to illustrate this phenomenon, with $Q$ serving as an approximate stability criterion, and $\cal T$ roughly expressing the efficiency of angular momentum and mass redistribution by spiral inhomogeneities within the disk \citep{VB06}. As we are interested in the global properties of the disk (instead of their local variations), we calculate an approximate global value of $Q=\tilde{c}_{\rm s} \Omega/(\pi G \Sigma)$, averaging $\tilde{c}_{\rm s}$, $\Omega$, and $\Sigma$  over all of the computational grid zones of the inner 500~AU region within which the disk is localized. The quantity $\tilde{c}_{\rm s}=(dP/d\Sigma)^{1/2}$ is the effective sound speed.

\begin{figure}
  \centering
  \includegraphics[width=\columnwidth]{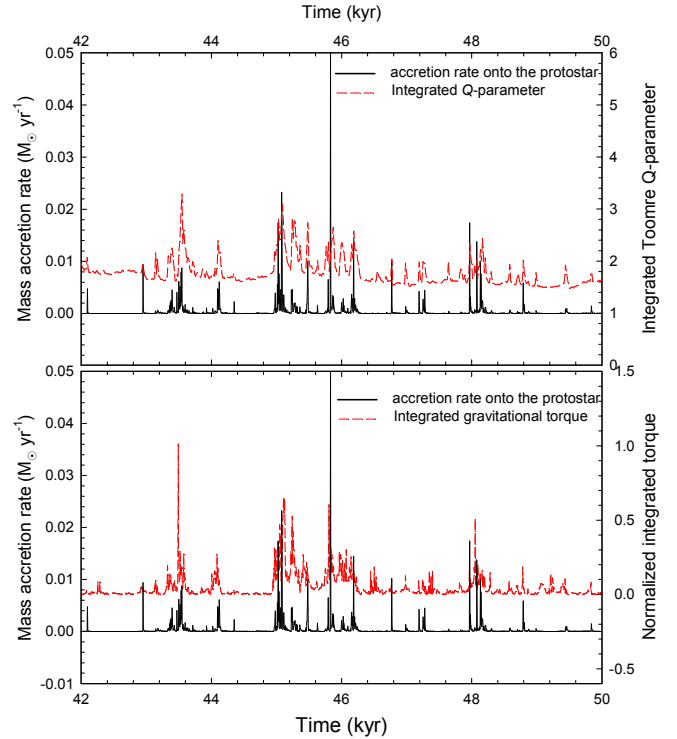}
  \caption{\textbf{Top:} Mass accretion rate onto the protostar (black solid line) and the global Toomre $Q$-parameter (red dashed line) versus time elapsed since the formation of the protostar. \textbf{Bottom:} Mass accretion rate onto the protostar (black solid line) and the integrated gravitational torque by absolute value (red dashed line) versus time. See text for an explanation of the interdependence between these quantities.}
  \label{fig10}
\end{figure}

Figure~\ref{fig10} presents the mass accretion rate onto the protostar (in black) and the integrated gravitational torque normalized to a local maximum value (in red) as a function of the time elapsed since the formation of the protostar from our reference model. We focus on a short period of the evolution in order to make the interdependence between the bursts, integrated torque, and $Q$-parameter clearly visible. For the same reason, we also plot the mass accretion rate on a linear scale here. The red line shows that $Q$ is smallest just before a burst, and its value rises sharply afterward, reflecting a transient decrease in the disk mass caused by fragments passing through the inner computational boundary. The integrated torque grows before each burst, reaching a maximum value at the moment of the burst itself due to strong gravitational interaction between spiral arms and fragments as the latter are torqued onto the protostar. After the burst, the integrated torque returns to a marginal value. We note that the $Q$-parameter in Figure~\ref{fig10} does not explicitly drop below unity---a typical value for gravitational fragmentation---because of the averaging over the entire disk; local values of $Q$ (Figure~\ref{fig4}) do drop below unity in the densest parts of the spiral arms and in the fragments.

\subsection{Parameter space study}

We next consider two additional models with the purpose of exploring the robustness of the accretion and luminosity burst phenomena. We choose primordial cores with a lower initial angular momentum (Model 1) and another with a higher initial temperature (Model 2) than in our reference model. These choices are motivated by numerical studies of the burst phenomenon in present-day star formation that have shown both the number and intensity of such bursts to diminish with decreasing angular momentum and increasing temperature of the parent core \citep{VB10}. The specific parameters of each model are listed in Table~\ref{table2}.

\begin{figure}
  \centering
  \resizebox{\hsize}{!}{\includegraphics{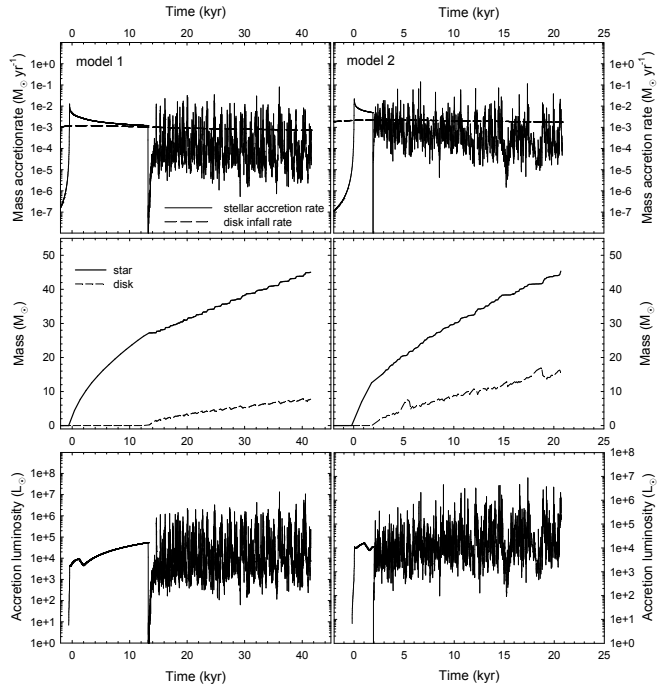}}
  \caption{Mass accretion rates (top row), disk and stellar masses (middle row), and accretion luminosity (bottom row) versus time elapsed after the formation of the protostar for Model 1 (left column) and Model 2 (right column). The solid lines in the top row show the mass accretion rate onto the protostar, while the dashed lines yield the mass infall rate at 3000~AU. The solid and dashed lines in the middle row are the masses of the protostar and the disk, respectively.}
  \label{fig11}
\end{figure}

Figure~\ref{fig11} presents the mass accretion rates (top row), disk and stellar masses (middle row), and accretion luminosity (bottom row) for Model 1 (left column) and Model 2 (right column). We terminate the simulations when the mass of the central objects reaches $45~\mbox{M}_\odot$. Although the burst phenomenon is still occurrent, the strength of the accretion and luminosity bursts is somewhat lowered in comparison with our reference model. Though some of the luminosity bursts exceed $10^7~\mbox{L}_\odot$ in the reference model, in Models 1 and 2 most of the bursts stay between $10^6$--$10^7~\mbox{L}_\odot$.

The apparent decline in the strength of the bursts in Model 1 is caused by a notable decrease in the corresponding disk mass as compared to the reference model. Model 1 is characterized by a factor of four smaller ratio of the rotational to gravitational energy $\beta$ than in the reference model, which leads to a delayed formation of the disk and overall lower disk mass. Low mass disks are less prone to gravitational fragmentation than their high mass counterparts, and the masses of their masses are correspondingly lower. The solid line in the middle-left panel in Figure~\ref{fig11} indicates that the typical mass of the accreted fragments in Model 1 does not exceed $1.0~\mbox{M}_\odot$, whereas in the reference model the fragments are of a few solar masses by the time they are accreted onto the protostar. This explains the somewhat weaker strength of the bursts in Model 1.

Model 2 has a higher initial temperature ($T=350$~K) and hence a higher disk temperature (because of the adopted polytropic equation of state) than the reference model ($T=250$~K). For instance, the gas temperature at $r=100$~AU in model~2 at $t=30$~kyr is approximately 1500~K, while in the reference model it is $\approx1100$~K. The contrast becomes stronger closer to the protostar. High temperature disks are also less prone to gravitational fragmentation because they have a higher Toomre $Q$-parameter than their low temperature counterparts. This effect is somewhat offset by a higher mass infall rate onto the disk as indicated by the dashed lines in Figure~\ref{fig11}. As a result, the variations in accretion and luminosity in Model 2 are of smaller amplitude than in the reference model, but the burst phenomenon is still well pronounced. We also ran a model with a factor of 1.5 smaller initial core  mass than that of the reference model but similar $\beta$, which yielded essentially similar results to those of Model 1. Obviously, models with increased $\beta$ and/or $M_{\rm c}$ that would yield an even more intense profile of bursts are not considered here. We conclude that the accretion and luminosity bursts are a robust phenomenon, which is expected to occur for a range of initial primordial core masses, angular momenta and temperatures.

\section{Discussion}

Our parameter study successfully demonstrates the robustness of the burst accretion phenomenon across a wide range of configurations in mass, rotation rates, and temperatures.
We find the formation of a quasi-Keplerian disk to be the outcome of the gravitational collapse of primordial prestellar cores. The disk expands quickly in radial extent to several hundred AU within just a few kyr of the formation of the central protostar. In agreement with \citet{Clark2011}, as has been found in studies of present day star formation \citep[e.g.,][]{VB2007}, we find that the disk exhibits a near-constant Toomre $Q$ parameter ${\sim}1$. Figure~\ref{fig11} demonstrates that this is not only true initially, but holds over a substantial tract of time, ${\sim}80~\mbox{kyr}$.

Studies of mass accretion during primordial star formation have focused primarily on the smooth accretion of material onto the protostar, reminiscent of Larson-Penston type analytic solutions \citep[e.g.,][]{OP2003}. Recently, \citet{Smith2011} studied the multiplicity of fragmentation events in somewhat larger cloud cores. They found the accretion rates varied due to the motions of the protostellar fragments through regions of alternately high- and low-density gas within the natal environment, with initial accretion rates of $10^{-2}~\mbox{M}_{\odot}~\mbox{yr}^{-1}$ that then declined to $10^{-3}~\mbox{M}_{\odot}~\mbox{yr}^{-1}$; in approximate agreement with analytic solutions. Indeed, we find similar mass accretion rates onto the disk of ${\sim}10^{-3}~\mbox{M}_{\odot}~\mbox{yr}^{-1}$. However, this source of infalling material is responsible for the mass loading of the disk and its subsequent fragmentation (Section 3.1), which is the ultimate origin of the variability in the actual rate of accretion onto the protostar.

The burst mode of accretion thus represents an entirely novel mechanism in contrast to either the classical analytic scenario of monolithic accretion, or that due to the motion of individual protostellar fragments. The resupply of material infalling onto the disk at large radii establishes a pattern of recurrent gravitational instability driven fragmentation, as has been seen in models of present-day disk evolution \citep[e.g.,][]{VB06}. In analogue studies by \citet{Smith2012} it has been noted that the number of fragmentation events actually increases steadily in time. We find a similar steady increase initially (Fig.~\ref{fig7}). However, SPH implementations may be limited in their ability to resolve small-scale torques that may act to shear objects apart, and long-term temporal resolution is required to resolve the variability and seeming periodicity in the rate of fragment formation observed in the simulations presented herein.

The large-amplitude variations in the mass accretion rates are also responsible for giving rise to a correspondingly large-amplitude time varying accretion luminosity (Fig.~\ref{fig8}). Although the accretion luminosity associated with primordial star formation is often regarded as only a mild heating source \citep{MT2008,Hosokawa2011}, the nature of episodic accretion suggests that the accretion luminosity itself may exceed prior estimates by a factor of between 10--100 times. This may raise the profile of the heating rate due to accretion to be comparable to that from gas compression. However, our polytropic modeling of the gas thermodynamics is the primary caveat limiting our ability to analyze the complete effect of this heating mechanism on the disk's evolution. The robustness of our simulation results, with respect to higher disk temperatures however, suggest that this mechanism may be self-regulating, with increased heating due to bursts stymieing fragmentation, allowing the disk to cool, and in-turn increasing again the fragmentation likelihood.

Although most fragments are torqued inward and onto the central protostar, contributing to the luminosity bursts, several alternative fates also exist. Fragments capable of settling into stable orbits may evolve in conjunction with their hosts to form binary-pairs of first stars \citep[e.g.,][]{Machida2008}. In fact, \citet{Suda2004} has shown that the abundance patterns of at least two hyper metal-poor ([Fe/H]$<$-5) stars \citep{Christlieb2001,Frebel2005} can be explained by a unique history of mass transfer onto a first-generation low-mass binary star.

Finally, we note that several of the fragments (F2 for example, in Figs.~\ref{fig4} and \ref{fig8}) are seen being ejected to substantially larger orbits as a result of N-body-like interactions. Recent studies by \citet{Clark2011} and \citet{Greif2011} noted similar dynamically chaotic interactions between fragments formed in those simulations. Although these fragments are most often sheared apart in the outer disk, we posit that a low-mass fragment---that is itself self-gravitating---may be ejected from the protostellar embryo as seen in simulations of present-day star formation \citep{VB2012}. This raises the tantalizingly possibility of a heretofore unseen population of low-mass primordial stars that may have survived into the present-day.

\section{Model caveats}

{\it The thin-disk approximation.} The applicability of the thin-disk approximation in our models is discussed in the Appendix. Here, we want to stress three additional points. First, the aspect ratio $A$ of the disk scale height $Z$ to radial distance $r$ strongly depend on the disk-to-star mass ratio $\xi$ (see equation~\ref{aspect}) and consequently on the initial conditions in the primordial cores. Massive cores with high angular momentum are expected to form massive disks soon after the formation of the protostar. These disks can be characterized by high $\xi \simeq 1.0$ and, consequently, by high $A\simeq1.0$ \citep{Clark2011}. Our models have $\xi<0.5$, which allows us to use the thin-disk limit. Second, the disk structure is quite irregular and the thin-disk approximation may break locally, even though it is fulfilled globally. And finally, full three-dimensional numerical simulations of present-day star formation \citep[e.g.][]{Machida2011} confirmed robustness of the burst phenomenon originally discovered using the thin-disk simulations \citep{VB05,VB06}. Three-dimensional simulations of primordial star formation already showed quick inward migration of the fragments on short timescales \citep{Greif2012} and we await confirmation of the repetitive nature of the burst phenomenon on long timescales.

{\it Barotropic equation of state.} In the present study, we approximated the thermal balance of the gas using a barotropic equation of state. During the collapse of a cloud core the major heating source comes from compression and the barotropic approximation is justified. However, once star formation is underway, radiative cooling and stellar irradiation may produce a range of temperatures that cannot be explained by a simple barotropic approximation \citep{Clark2011b}.
Numerical simulations of fragmenting barotropic disks seem to yield more fragments than those that take into account a detailed thermal balance \citep[e.g.][]{Stamatellos09}. We note, however, that the burst phenomenon in the present-day star formation, originally discovered using a barotropic equation of state \citep{VB06}, was later shown to exist when more detailed thermal physics calculations were taken into account \citep{VB10}.\\

{\it Stellar irradiation.} A burgeoning protostar may start affecting its surroundings quite early in the evolution.
For the mean accretion rate of $10^{-3}~M_\odot$~yr$^{-1}$, a characteristic upper limit in our models, stellar UV irradiation becomes notable at a mass of about $M_\ast=15~M_\sun$ \citep{Smith2011,Hosokawa2012}. Our numerical simulations extend to the time instant when the mass of the protostar reaches $45~M_\odot$, meaning that the disk evolution may be affected by the lack of stellar irradiation. Numerical simulations taking into account a more detailed thermal balance are needed to access the effect of stellar irradiation on the burst phenomenon.  \\

\section{Conclusions}

We have investigated the gravitationally-induced collapse of prestellar cores having pristine primordial gas composition, using nonaxisymmetric hydrodynamics simulations in the thin-disk limit. The gas thermal chemistry is modelled with a barotropic relation adapted from \citet{OP2003}. We follow these simulations, in the absence of protostellar feedback, to the point at which UV ionizing irradiation from the central star would become important (while $M\,{\la}\,45~\mbox{M}_{\odot}$). Our main conclusions are as follows:

\begin{itemize}
\item Recurrent gravitational instability driven fragmentation and accretion of the fragments is an important mechanism through which protostars accumulate mass in the early universe. This mechanism is mediated by smooth mass infall from the surrounding envelope at $\dot{M}\,{\approx}\,10^{-3}~\mbox{M}_{\odot}~\mbox{yr}^{-1}$. As mass is loaded onto the disk, gravitational instability eventually induces fragments to form, and these are then driven onto the protostar---resulting in $\dot{M}$ of up to $10^{-1}~\mbox{M}_{\odot}~\mbox{yr}^{-1}$---by gravitational torques acting within the disk.
\item The burst mode of accretion is sensitive to variations in the initial conditions (mass, rotation rate, and temperature) of the parent core, with the strength of bursts decreasing for decreasing $\beta$---the ratio of the magnitudes of rotational to gravitational energy---and/or increasing disk temperature $T$.
\item We have considered cloud core masses and rotation rates that are somewhat lower than those derived from the numerical hydrodynamics simulations of collapsing primordial mini-halos \citep[e.g.][]{Yoshida06,Clark2011}. Even in this case however, the burst phenomenon could not be entirely suppressed, and is expected to be more vigourous for other possible initial conditions. We conclude that the burst mode of accretion is likely a robust phenomenon in primordial star formation.
\item Accretion luminosity produced by episodic mass accretion may contribute significantly to the heating of material in the immediate vicinity of the protostar, but numerical simulations taking into account the heating/cooling balance are needed to assess this effect. 
\item Not all fragments eventually migrate inward and onto the protostar; some of the fragments are seen being ejected to the outer regions of the disk, in line with previous studies by \citet{Clark2011} and \citet{Greif2011}. Given the implications for low-mass primordial star formation, the likelihood for survival of these fragments needs to be studied. 
\end{itemize}

\acknowledgements
We are grateful to the referee for useful suggestions and comments that helped to improve the paper. The authors thank M.~N.~Machida for providing data after which the density-temperature relation discussed herein was modeled. E.~I.~V.~acknowledges partial support from the RFBR grant 10-02-00278. S. B. acknowledges support from an NSERC Discovery Grant. The simulations were performed on the Shared Hierarchical Academic Research Computing Network (SHARCNET), on the Atlantic Computational Excellence Network (ACEnet), and on the Vienna Scientific Cluster (VSC-2).

\appendix
\section{Thin-disk approximation}
The applicability of the thin-disk approximation hinges upon the geometry of the considered configurations. Elongated shapes are typical for protostellar disks in nearby star-forming regions.  Specifically, the thin-disk approximation is well justified as long as the aspect ratio $A = Z / r$ of the disk vertical scale height $Z$ to radial distance in the plane of the disk $r$ does not considerably exceed 0.1. This condition is usually fulfilled for disks with sizes up to 1000~AU \citep[see e.g.,][]{VB10}, but its applicability may be less obvious for the evolution phase immediately preceding disk formation, when material accretes directly from the infalling core onto the forming star. Nevertheless, the temporal behaviour and magnitude of the accretion rate onto the star is very similar in the pre-disk stage for spherically symmetric and disk-like cores \citep{VB05,VB06}, ensuring that the stellar masses are calculated accurately in our simulations. Below we provide some analytical and numerical estimates justifying our use of the thin-disk approximation.

\begin{figure}
  \centering
  \includegraphics[width=8cm]{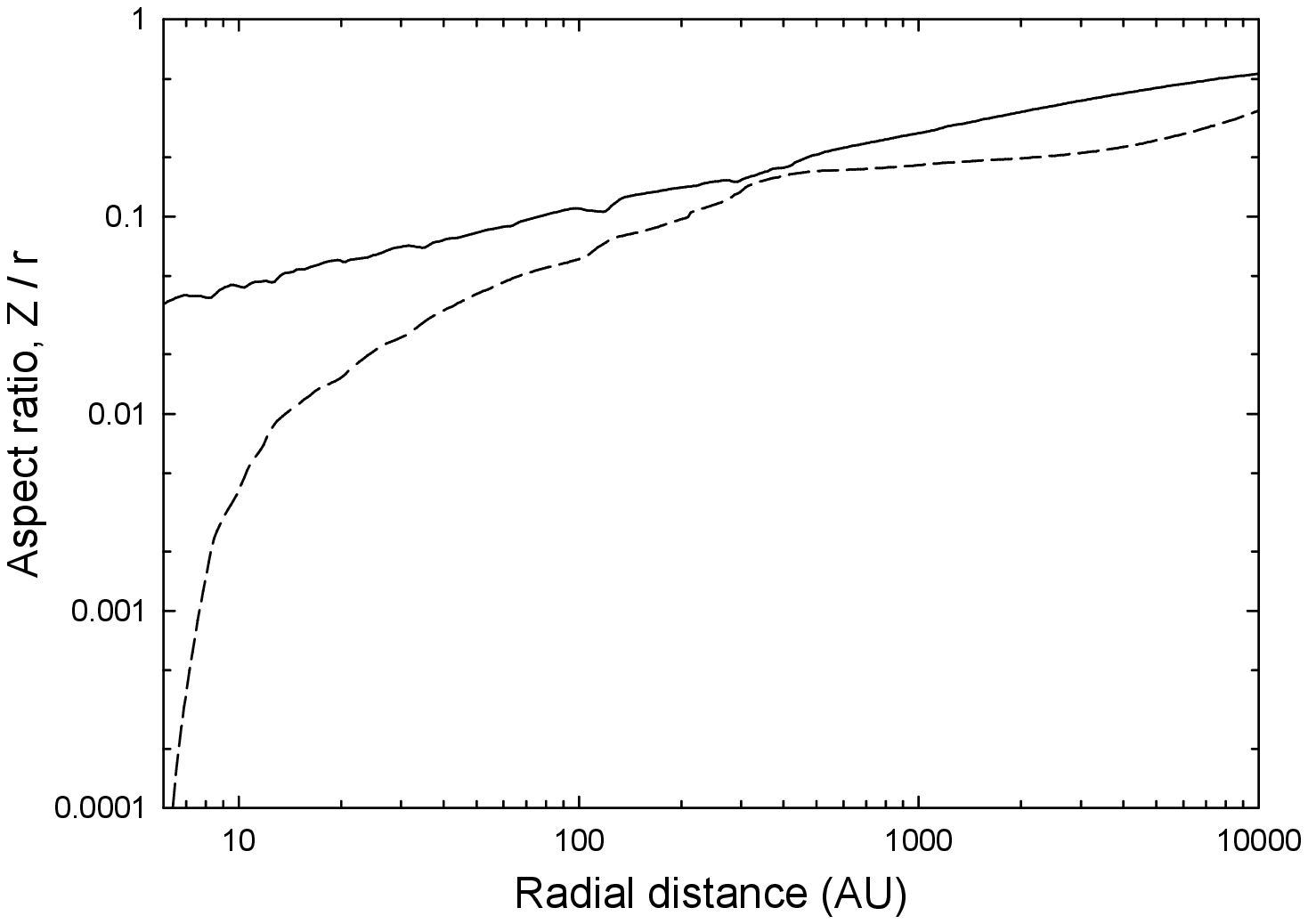}
  \caption{Aspect ratio $A$ of the disk vertical scale height to radius ($Z/r$) as a function of radius ($r$) in the reference model at $t=30$~kyr after the formation of the central star. The solid line presents
the data calculated using the assumption of local vertical hydrostatic equilibrium and the dashed line
is derived using equation~(\ref{aspect}).}
  \label{fig12}
\end{figure}

In a Keplerian disk, the aspect ratio $A=Z/r$ can be expressed as \citep{VB10}
\begin{equation}
A \le { Q_{\rm cr} \, M_{\rm d}(r) \over C M_\ast},
\label{aspect}
\end{equation}
where $M_{\rm d}(r)=\int\Sigma(r,\phi)\, r \,dr \,d\phi$ is the disk mass contained within radius $r$, $M_\ast$ is the mass of the central star, $Q_{\rm cr}$ is the critical Toomre parameter, and $C$ is
a constant, the actual value of which depends on the gas surface density distribution $\Sigma$ in 
the disk. For a disk of constant surface density, $C$ is equal to unity and for the 
$\Sigma \propto r^{-1.5}$ scaling  typical for gravitationally unstable disks (see Fig.~\ref{fig7}),
$C=4$.
Adopting a conservative value of $C=2$ and setting $Q_{\rm cr}$ to 
unity---characteristic for fragmenting disks---we obtain the radial profile of $A$ shown 
in  Figure~\ref{fig12} by the dashed line. The ratio $M_{\rm d}(r)/M_\ast$ was calculated using
the reference model at $t=30$~kyr.
Evidently, the aspect ratio derived using analytic considerations is smaller than 0.2 in the inner several
thousand AU, which validates our use of the thin-disk approximation. 
We note however that this approximation may become only marginally valid at large $r$ where 
$M_{\rm d}(r)/M_\ast$ approaches its maximum value.

This estimate is confirmed by the exact calculations of the disk scale height in our models. The azimuthally averaged radial distribution of the aspect ratio $A = Z/r$ in the reference model at $t=30~\mbox{kyr}$ after the formation of the central star is shown by the solid line in Figure~\ref{fig12}. The vertical scale height $Z$ is calculated assuming local vertical hydrostatic equilibrium in the disk using the method described in \citet{VB09}. Figure~\ref{fig12} reinforces our analytical estimate and demonstrates that the thin disk approximation is certainly obeyed within the actual disk, which, according to Figures~\ref{fig4} and \ref{fig7}, does not exceed 1000~AU in radius. Within this radial extent, the corresponding aspect ratio is below 0.25. Only at radial distances well in excess of 1000 AU might the thin-disk approximation be violated.

Finally, we note that the thin-disk approximation assumes the absence of vertical motions, which
turns the usual momentum equation for the $z$-component of the gas velocity into the equation describing
the local vertical hydrostatic equilibrium.
This assumption is used to calculate the scale height $Z$ and the aspect ratio $A$ in equation~(\ref{aspect}).
Therefore, a possible thick disk that is not in vertical hydrostatic equilibrium is not accessible through our modeling.


\section{Fragment identification algorithm}
In the absence of sink particles, fragment identification on the computational mesh becomes a challenging task. Although from a numerical point of view the fragments are no different than from the rest of the disk, they can be identified based on the following set of physical criteria. First, we scan the disk and locate local maxima in the gas surface density that satisfy
\begin{equation}
\label{critdens}
\Sigma > \Sigma_{100} \left( {r \over 100~\mathrm{AU}} \right)^{-1.5},
\end{equation}
where $\Sigma_{100}$ is the gas surface density at a distance of 100~AU from the central star, and $r$ is the radial distance in AU. A value of $\Sigma_{100}=5000~\mbox{g~cm}^{-2}$ is chosen to represent typical densities of the fragments at 100~AU, based on the radial gas surface density profiles shown in Figure~\ref{fig7}. Consequently, this criterion helps to filter out local maxima, e.g., spiral arms, while retaining those that represent the true fragments, which are usually characterized by a much higher density than the rest of the disk.

After the radial and angular coordinates ($r_{\rm c}$, $\phi_{\rm c}$) of the local maximum representing the center of a fragment have been identified on the computational mesh, we determine which of the neighbouring cells also belong to the fragment by imposing the following two conditions on the gas pressure $P$ and gravitational potential $\Phi$
\begin{eqnarray}
\label{pres}
{\partial P \over \partial r^\prime} &+& {1 \over r^\prime}{\partial P \over \partial \phi^\prime} <0, \\
\label{grav}
{\partial \Phi \over \partial r^\prime} &+& {1 \over r^\prime}{\partial \Phi \over \partial \phi^\prime} >0,
\end{eqnarray}
where $r^\prime=r-r_{\rm c}$ and $\phi^\prime=\phi-\phi_{\rm c}$. The first condition mandates that the fragment must be pressure supported, with a negative pressure gradient with respect to the center of the fragment. The second condition requires that the fragment be kept together by gravity, with the potential well being deepest at the center of the fragment. Although substantial support against gravity may be provided by rotation, we assume it to not invalidate the first criterion.

In practice, we start from the center of the fragment and proceed in eight directions (along the coordinate directions and also at a median angle to them) until one of the above gradient criteria (or both) are violated. This procedure helps to outline an approximate shape of the fragment. We then check all the remaining cells that are encompassed by this shape and retain only those that meet both criteria. In addition, we filter out those cells with a gas surface density lower than that defined by equation~(\ref{critdens}) with $\Sigma_{100}=5000~\mbox{g~cm}^{-2}$.

\end{document}